\documentclass[twocolumn,showpacs,preprintnumbers,amsmath,amssymb]{revtex4}

\usepackage{graphicx}
\usepackage{dcolumn}
\usepackage{bm}

\begin{document}

\preprint{APS/123-QED}


\title{Absolute Negative Conductivity
in Two-Dimensional Electron Systems \\ 
Associated with
Acoustic Scattering Stimulated by Microwave Radiation
} 
\author{ 
V.~Ryzhii }
\email{v-ryzhii@u-aizu.ac.jp}
\author{V.~Vyurkov}
\affiliation{Computer Solid State Physics Laboratory, University of Aizu,
Aizu-Wakamatsu 965-8580, Japan}


\date{\today}

\begin{abstract}
We discuss the feasibility  of absolute negative conductivity (ANC)
in two-dimensional electron systems 
(2DES)
stimulated by microwave radiation in transverse magnetic field.
The mechanism of ANC under consideration
is associated with the electron scattering on acoustic piezoelectric
 phonons
accompanied by the absorption of microwave photons.
It is demonstrated that the  dissipative components of the 2DES
dc conductivity can be negative ($\sigma_{xx} = \sigma_{yy} < 0$)
when the microwave frequency $\Omega$ is somewhat higher  than
the electron
cyclotron frequency $\Omega_c$ or its harmonics.
The concept of ANC associated with such a scattering mechanism 
can be invoked to  explain 
the nature of the occurrence of  zero-resistance ``dissipationless'' 
states observed in recent experiments.
\end{abstract}

\pacs{PACS numbers: 73.40.-c, 78.67.-n, 73.43.-f}


\maketitle

\section{Introduction}

The effect of vanishing electrical resistance 
and transition to ``dissipationless'' states in a two-dimensional
electron system (2DES) subjected to a  magnetic field and
irradiated with microwaves
has recently been observed in the experiments by Mani {\it et al.}~\cite{1}, 
Zudov {\it et al}.~\cite{2}, and Yang {\it et al}~\cite{3}. 
The most popular scenario~\cite{4,5,6} 
of the occurrence of this effect
is based on the concept of absolute negative conductivity (ANC)
in 2DES  proposed more than three decades ago by Ryzhii~\cite{7}.
Later~\cite{8}, a detailed theory of the ANC effect 
in 2DES due to the photon-assisted
impurity scattering 
was developed. Recently, a new version of the theory of
this effect was
presented  by Durst {\it et al.}~\cite{9}.
Some nonequilibrium processes in 2DES associated with the
electron-phonon interactions were studied theoretically 
as well~\cite{10,11,12,13}.

The feasibility of ANC in heterostructures with a 2DES
in the magnetic field under microwave irradiation is associated
with the following. The dissipative electron transport in the direction
parallel to  the electric field and perpendicular to the magnetic field
is due to hops of the electron Larmor orbit centers caused by
scattering processes. These hops result in a change in the electron
potential energy $\delta \epsilon = eE\delta\xi$,
where $e = |e|$ is the electron charge,
$E = |{\bf E}|$ is the modulus of the net in-plane
electric field, which includes both the applied and
the Hall components, and $\delta\xi$ is the displacement of 
the electron Larmor orbit center. If the electron Larmor orbit center
shifts in the direction opposite to the electric field ($\delta\xi < 0$),
the electron potential energy decreases $(\delta \epsilon < 0$).
In equilibrium, the electron Larmor orbit center hops in this
direction dominate, so the dissipative electron current flows in
the direction of the electric field. However, in some cases,
the displacements of the electron Larmor orbit centers in the direction
of the electric field(with  $\delta \epsilon > 0$) can prevail.
Indeed, under sufficiently strong
microwave irradiation the main contribution to the electron
scattering on impurities can be associated with the processes
involving the absorption of
the microwave photons. If an electron absorbs such a photon
and transfers to a higher Landau level (LL), a portion of the
absorbed energy $N\hbar \Omega_c$, where  $\Omega_c = eH/mc$
is the electron cyclotron frequency, $\hbar$ is the Planck constant,
 $m$ is the electron effective mass, $H$ is the strength of the magnetic
field, $c$ is the velocity of light, and $N = 1, 2, 3, ...$
is the LL index,
goes to an increase
of the electron kinetic energy, whereas the change
in the electron potential energy
is
$\delta \epsilon = \hbar(\Omega - \Lambda \Omega_c)$, where 
$\Lambda = N^{\prime} - N$. 
If $(\Omega - \Lambda\hbar \Omega_c) > 0$, the potential energy of electrons
increases with each act of their scattering. Hence, the dissipation
current flows opposite to the electric filed resulting in
negative (absolute, not differential) conductivity. 
However, the probability of the scattering with
the spatial displacements of the electron Larmor orbit center  $\xi$
exceeding the quantum Larmor radius $L = (c\hbar/eH)^{1/2}$
 is exponentially small.
Due to this, such scattering processes are effective,
and the variation of the dissipative component of the current
caused by microwave radiation (i.e., the photocurrent)
is be substantial only 
in the immediate vicinities of the resonances
 $|\Omega - \Lambda\Omega_c| 
\lesssim max\{eEL/\hbar,\,\Gamma\}$,
where $\Gamma$ characterizes the LL broadening~\cite{14}.
Hence, at small $eEL/\hbar$ and $\Gamma$,
the ranges  $\Omega - \Lambda\Omega_c$ in which  ANC associated with
the photon-assisted impurity scattering  occurs
are rather narrow.
Therefore, the contributions of  other scattering mechanisms should 
be assessed.
In this paper, we calculate the dc dissipative components
of the 2DES conductivity (mobility) tensor, assuming
that the main scattering mechanisms  are the electron scattering 
 and the electron photon-assisted
scattering on the piezoelectric acoustic  phonons.
This assumption is supported by
high quality of the samples (with the electron mobility
$\mu > 10^7$~cm$^2$/V\,s) used
in the experiments~\cite{1,2,3} performed at low temperatures ($T \simeq 1$K).

\section{General formulas}

The density of the in-plane  dc current in a 2DES in the transverse
magnetic field is given by the following equations:

\begin{equation}\label{eq1}
j_x = \sigma(E) E_x + \sigma_HE_y,\, j_y =  \sigma(E)E_y - \sigma_HE_x,
\end{equation}
where $\sigma(E)$ and $\sigma_H$ are 
the  in-plane dissipative and Hall  components
of the dc conductivity tensor, respectively,
 with  $\sigma(E) = \sigma_{xx} = \sigma_{yy}$
and, considering
the classical Hall effect, 
$\sigma_H = \sigma_{xy} = - \sigma_{yx} = ec\Sigma/H$.
Here,
$\Sigma$  is the electron sheet concentration,
and the directions $x$ and $y$ are in the 2DES plane. 
Positive values of  $\sigma(E)$ correspond to
the electron drift in the direction opposite to the direction
of the electric field, i.e., to the usual conductivity,
whereas the case $\sigma(E) < 0$ is referred to as ANC.

We shall assume that 
$\hbar \Omega,\, \hbar \Omega_c > T$,
where $T$ is the temperature in the energy units.
In this case, one can disregard the processes
accompanied by the 
emission of microwave photons. 

As follows from Eq.~(1), the density of the dissipative current
i.e., the current parallel to
the net electric field ${\bf E}$, can be presented 
as $j_D(E) = \sigma(E)E$.
The dissipative current is determined by the
transitions of  electrons between their quantum states
with the change in   the  coordinate of  
the electron Larmor orbit center $\xi$.
The energies of the electron states in a 2DES in crossed magnetic and electric
fields (neglecting the Zeeman splitting) are given by
\begin{equation}\label{eq2}
\epsilon_{N,\xi} = \biggl(N + \frac{1}{2}\biggr)\hbar\Omega_c
+ eE\xi.
\end{equation}

Adjusting the standard ''orthodox'' approach~\cite{15} for 
the calculation of the dissipative current in a
2DES  (see, for example, Refs.~7,10,12,16,17),
$j_D(E) $ can be presented in the form
$$
j_D(E) = \frac{e}{\hbar} \sum_{N,N^{\prime} }
f_N(1 - f_{N^{\prime}})
$$
$$
\times\int d^2{\bf q}\,
q_y|V_{{\bf q}}|^2|Q_{N,N^{\prime}}(L^2q^2_{\perp}/2)|^2
$$
$$
\times\{
{\cal N}_q
\delta[(N - N^{\prime})\hbar\Omega_c + \hbar\omega_q + eEL^2q_y]
$$
$$
+ ({\cal N}_q + 1)
\delta[(N - N^{\prime})\hbar\Omega_c - \hbar\omega_q + eEL^2q_y] 
$$
$$
+ I_{\Omega}(q_x,q_y){\cal N}_q
\delta[\hbar\Omega + (N - N^{\prime})\hbar\Omega_c + \hbar\omega_q + eEL^2q_y]
$$
\begin{equation}\label{eq3}
+  I_{\Omega}(q_x,q_y)({\cal N}_q + 1)
\delta[\hbar\Omega + (N - N^{\prime})\hbar\Omega_c - \hbar\omega_q + eEL^2q_y]\}.
\end{equation}
Here,
$f_N$ is the filling factor of the $N$th Landau level given
by the Fermi distribution function,
${\bf q} = (q_x, q_y, q_z)$, $\omega_q = sq$, and  
${\cal N}_q = [\exp(\hbar\omega_q/T)  - 1]^{-1}$ are 
the phonon wave vector, frequency, and   
distribution function, respectively, $s$ is the velocity of sound,
$q = \sqrt{q_x^2 + q_y^2 + q_z^2}$, 
$q_{\perp} =  \sqrt{q_x^2 + q_y^2}$, 
$\delta (\omega)$ is  the
LL form-factor which at small
$\Gamma$ can be assumed to be the Dirac delta function,
$|V_{{\bf q}}|^2 \propto q^{-1}\exp(- l^2q_z^2/2)$ characterizes
the piezoelectric interaction of electrons with acoustic  phonons
(because such an interaction is considered as
most important in the 2DES under consideration
at low temperatures~\cite{18} ), $l$ is
the width of the electron localization in the $z$-direction perpendicular to
the 2DES plane ($l \ll L$), and
$|Q_{N,N^{\prime}}(L^2q^2_{\perp}/2)|^2  = |P_N^{N^{\prime} - N}(L^2q^2_{\perp}/2)|^2 \exp (- L^2q^2_{\perp}/2))$ 
is the matrix
element determined by the overlap of the electron wave functions
before and after the hop caused by the scattering,
$|P_N^{N^{\prime} - N}(L^2q^2_{\perp}/2)|^2$
is proportional to a Laguerre  polynomial.

The quantity  $I_{\Omega}(q_x,q_y)$ is proportional to
the incident microwave power.
It characterizes the effect
of microwave field on the in-plain electron motion.
For the nonpolarized microwave radiation~\cite{19},
$I_{\Omega}(q_x,q_y) = {\cal J}_{\Omega}L^2(q_x^2 + q_y^2)$, where
${\cal J}_{\Omega} = ({\cal E}_{\Omega}/\tilde{{\cal E}}_{\Omega})^2$,
${\cal E}_{\Omega}$ is the microwave electric field amplitude,
which is assumed to be smaller than
some characteristic microwave field 
$\tilde{{\cal E}_{\Omega}}$.
The latter assumption implies that 
the Larmor orbit center oscillation amplitude in the microwave
field is smaller than $L$,  
and that  the radiative
processes with the participation
of more than one microwave photon are insignificant.
 The inclusion of the polarization effects leads
to the appearance of some anisotropy at the frequencies
far from  the cyclotron resonance.
A more general formula  can be used in line with Ref.~19
if ${\cal J}_{\Omega}$ is of the order of unity or larger.  
Deriving Eq.~(3), we have taken into account
that the displacement of the electron Larmor orbit center is
 $\delta\xi = - L^2q_y$.
First two terms in the right-hand side of Eq.~(3)
correspond to the electron-phonon interactions whereas
the third and forth terms are associated with such interactions
accompanied by the absorption of a photon.
Worth pointing out that, as can be seen from Eq.~(3), 
at the resonances $\Omega = (N^{\prime} - N)\Omega_c$,
the contribution of the photon-assisted processes to
the dissipative conductivity turns  zero (see, Refs.~\cite{7,8}).
The characteristic
amplitude can be presented as~\cite{18}
\begin{equation}\label{eq4}
\tilde{{\cal E}_{\Omega}} = \frac{\sqrt{2
}m\Omega|\Omega_c^2 - \Omega^2|L}
{e\sqrt{\Omega_c^2 +  \Omega^2}}.
\end{equation}
In the immediate vicinity of the cyclotron resonance $\Omega = \Omega_c$, 
the quantity $\tilde{{\cal E}_{\Omega}}$ is limited by the LL broadening,
Eq.~(4) becomes invalid, and 
$\tilde{{\cal E}_{\Omega}}$ 
can be estimated as $\tilde{{\cal E}_{\Omega}} 
\simeq \sqrt{2}m\Omega\Gamma/e$.

Assuming that 
$eE|\xi| \ll \hbar|\Omega - \Lambda\Omega_c| <  \hbar \Omega, 
\hbar \Omega_c$, one can expand  the expression for the dissipating current
given by Eq.~(3) in powers of $(eEL^2/\hbar s)$ and present 
the dissipative dc conductivity in the following form:
\begin{equation}\label{eq5}
\sigma(E)  \simeq \sigma_{dark} + 
\sigma_{ph},
\end{equation}
in which the terms of the order of $(eEL^2/\hbar s)^3$ 
and higher have been neglected. 
As a result, 
from Eq.~(3), we obtain
$$
\sigma_{dark} = \biggl(\frac{e^2L^2}{\hbar^3s^2}\biggr) 
\sum_{N,\Lambda}
\int d^3{\bf q}\,
q_y^2|V_{{\bf q}}|^2
$$
$$
\times|Q_{N,N+\Lambda}(L^2q^2_{\perp}/2)|^2
\delta^{\prime}(q  - q^{\Lambda})
$$
\begin{equation}\label{eq6}
\times\{[f_N(1 - f_{N+\Lambda}){\cal N}_q - f_{N+\Lambda}(1 - f_N)({\cal N}_q + 1)],
\end{equation}
$$
\sigma_{ph} = {\cal J}_{\Omega}\biggl(\frac{e^2L^4}{\hbar^3s^2}\biggr)  
\sum_{N,\Lambda>0}
\int d^3{\bf q}\,
q_y^2q^2_{\perp}|V_{{\bf q}}|^2
$$
$$
\times|Q_{N,N+\Lambda}(L^2q^2_{\perp}/2)|^2
\biggl[
{\cal N}_q\delta^{\prime}(q  + q_{\Omega}^{\Lambda})
$$
\begin{equation}\label{eq7}
- ({\cal N}_q + 1)
\delta^{\prime}(q  - q_{\Omega}^{\Lambda})\biggr]
f_N(1 - f_{N+\Lambda}),
\end{equation}
where $\delta^{\prime}(q) = d\delta(q)/dq$ ,
$q^{(\Lambda)} = \Lambda\Omega_c/s$, $q_{\Omega}^{(\Lambda)}
 = (\Omega - \Lambda\Omega_c)/s$, and $\Lambda > 0$.
Substituting  the integration over $d^3{\bf q}$ for the integration
over $dq\,dq_{\perp}d\theta$, where $sin \theta = q_y/q_{\perp}$,
Eqs.~(6) and (7)
can be rewritten as 
$$
\sigma_{dark}  \propto 
\sum_{N\Lambda}
\int_0^{\infty}
\frac{dqdq_{\perp}q_{\perp}^3}
{\sqrt{q^2 - q_{\perp}^2}}\exp\biggl[- \frac{l^2(q^2 - q_{\perp}^2)}{2}\biggr]
$$
$$
\times|Q_{N,N+\Lambda}(L^2q^2_{\perp}/2)|^2\delta^{\prime}(q  - q^{(\Lambda)})
$$
\begin{equation}\label{eq8}
\times
[f_N(1 - f_{N+\Lambda}){\cal N}_q - f_{N+\Lambda}(1 - f_N)({\cal N}_q + 1)],
\end{equation}
$$
\sigma_{ph} \propto {\cal J}_{\Omega}L^2 
\sum_{N,\Lambda}
\int_0^{\infty}
\frac{dqdq_{\perp}q_{\perp}^5}
{\sqrt{q^2 - q_{\perp}^2}}\exp\biggl[- \frac{l^2(q^2 - q_{\perp}^2)}{2}\biggr]
$$
$$
\times
|Q_{N,N+\Lambda}(L^2q^2_{\perp}/2)|^2
\biggl[
{\cal N}_q\delta^{\prime}(q  + q_{\Omega}^{(\Lambda)})
$$
\begin{equation}\label{eq9}
- ({\cal N}_q + 1)
\delta^{\prime}(q  - q_{\Omega}^{(\Lambda)})\biggr]
f_N(1 - f_{N+\Lambda}),
\end{equation}
and reduced to

$$
\sigma_{dark}  \propto 
\sum_{N,\Lambda}f_N(1 - f_{N+\Lambda})
\int_0^{\infty}
dq 
\exp\biggl(- \frac{l^2q^2}{2}\biggr)
G_{N}^{\Lambda}(q)
$$
\begin{equation}\label{eq10}
\times
[{\cal N}_q - \exp( - \Lambda\hbar\Omega_c/T) ({\cal N}_q + 1)]\delta^{\prime}(q  - q^{(\Lambda)}),
\end{equation}

$$
\sigma_{ph}  \propto  {\cal J}_{\Omega}\sum_{N,\Lambda}
f_N(1 - f_{N+\Lambda})
\int_0^{\infty}
dq\exp\biggl(- \frac{l^2q^2}{2}\biggr)H_N^{(\Lambda)}(q)
$$
\begin{equation}\label{eq11}
\times
\biggl[
{\cal N}_q\delta^{\prime}(q  + q_{\Omega}^{(\Lambda)})
- ({\cal N}_q + 1)
\delta^{\prime}(q  - q_{\Omega}^{(\Lambda)})\biggr],
\end{equation}
where 
\begin{equation}\label{eq12}
G_N^{(\Lambda)}(q) = \int_0^{Lq} dtt^3
\frac{\displaystyle\exp\biggl[\frac{(l^2 - L^2)}{2L^2}t^2\biggr]}
{\sqrt{L^2q^2 - t^2}}
|P_N^{\Lambda}(t^2/2)|^2,
\end{equation}

\begin{equation}\label{eq13}
H_N^{(\Lambda)}(q) = \int_0^{Lq} dtt^5
\frac{\displaystyle\exp\biggl[\frac{(l^2 - L^2)}{2L^2}t^2\biggr]}
{\sqrt{L^2q^2 - t^2}}
|P_N^{\Lambda}(t^2/2)|^2.
\end{equation}
At $N \gg 1$,
using the asymptotic expressions for the Laguerre polynomials~\cite{20},
one can obtain $|P_N^{(\Lambda)}(t^2/2)|^2 \simeq J_\Lambda^2(\sqrt{2N_m}t)$,
where $J_\Lambda(t)$ is
the Bessel function,
so that $|P_N^{(\Lambda)}(t^2/2)|^2 \simeq
\cos^2[\sqrt{2N}t - (2\Lambda + 1)\pi/4]/\pi Nt$ if $t \gg 1/\sqrt{2N}$,
and $|P_N^{(\Lambda)}(t^2/2)|^2 \simeq t^{2\Lambda}$
when  $t < 1/\sqrt{2N} \ll 1$.
Therefore, 
$G_N^{(\Lambda)}(q)$ and $H_N^{(\Lambda)}(q)$ become 

$$
G_N^{(\Lambda)}(q) \simeq \frac{1}{\pi NLq}\int_0^{\infty} 
dt\,t^{2}
\exp\biggl[\frac{(l^2 - L^2)}{2L^2}t^2\biggr]
$$
$$
\times\cos^2\biggl[\sqrt{2N}t -\frac{(2\Lambda + 1)\pi}{4}\biggr],
$$
$$
H_N^{(\Lambda)}(q) \simeq \frac{1}{\pi NLq}\int_0^{\infty} 
dt\,t^{4}
\exp\biggl[\frac{(l^2 - L^2)}{2L^2}t^2\biggr]
$$
$$
\times\cos^2\biggl[\sqrt{2N}t -\frac{(2\Lambda + 1)\pi}{4}\biggr]
$$
at  $Lq \gg 1$, and 
$$
G_N^{(\Lambda)}(q) \simeq \frac{1}{\pi N}\int_0^{Lq} dt
\frac{t^{2}}
{\sqrt{L^2q^2 - t^2}}
\cos^2\biggl[\sqrt{2N}t -\frac{(2\Lambda + 1)\pi}{4}\biggr],
$$
$$
H_N^{(\Lambda)}(q) \simeq \frac{1}{\pi N}\int_0^{Lq} dt
\frac{t^{4}}
{\sqrt{L^2q^2 - t^2}}
\cos^2\biggl[\sqrt{2N}t -\frac{(2\Lambda + 1)\pi}{4}\biggr]
$$
at $1/\sqrt{2N} < Lq < 1$. 
After the integration involving averaging over
fast oscillations,
one can obtain
$
G_N^{(\Lambda)}(q) 
\simeq [2\sqrt{2\pi}(1 -l^2/L^2)^{3/2}NLq]^{-1}
\simeq (2\sqrt{2\pi}NLq)^{-1} $
at  $Lq~\gg~1$,
and 
$G_N^{(\Lambda)}(q) \simeq (Lq)^{2}/8 N$
in the range $1/\sqrt{2N} < Lq < 1$.
Analogously, for $H_N^{(\Lambda)}(q)$
in these ranges one obtains, respectively,
$H_N^{(\Lambda)}(q) 
\simeq 3[2\sqrt{2\pi}(1 -l^2/L^2)^{5/2}NLq]^{-1}
\simeq 3(2\sqrt{2\pi}NLq)^{-1} $
and 
$H_N^{(\Lambda)}(q) \simeq 3(Lq)^{4}/32 N$.
The derivative of
function $H_N^{(\Lambda)} (q)$ is proportional to
the average displacement of the electron Larmor orbit center
in the direction of the electric field
associated with the electron transition between the  $N$th and $(N + 1)$th
LL's involving  a microvave photon and an acoustic
phonon with the energies $\hbar\Omega$ and $\hbar s q$, respectively.
As an example, function $H_{50}^{(1)} (q)$
calculated using Eq.~(13), in which the asymptotic $|P_N^{(\Lambda)}(t^2/2)|^2 \simeq J_\Lambda^2(\sqrt{2N_m}t)$ was used,
is shown in Fig.~1.

\begin{figure}
\begin{center}
\includegraphics[width=7.5cm]{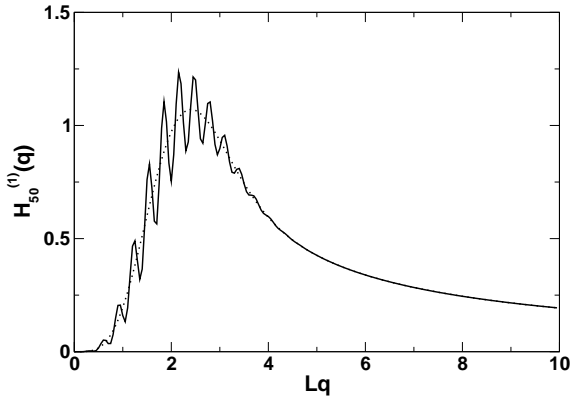}
\end{center}
\label{fig1}
\caption{Function $H_{50}^{(1)}(q)$ 
given by Eq.~(13). Dashed line corresponds to
averaging over fast oscillations.
}
\end{figure}

\section{Dark conductivity}

Considering the explicit formula for the phonon distribution (Planck's)
function, Eq.~(10)  can be presented in the following form:

$$
\sigma_{dark}  \propto 
\sum_{N,\Lambda > 0}
f_N(1 - f_{N+\Lambda})\int_0^{\infty}
dq\exp\biggl(- \frac{l^2q^2}{2}\biggr)G_N^{\Lambda}(Lq)
$$
\begin{equation}\label{eq14}
\times
\frac{1 - \exp[\hbar(sq - \Lambda\Omega_c)/T]}{\exp(\hbar sq/T) - 1}
\delta^{\prime}\biggl(q  - \frac{\Lambda\Omega_c}{s} \biggr),
\end{equation}
Neglecting
the terms containing higher 
powers of $\exp(- \hbar\Omega_c/k_BT)$ 
and denoting 
$$
G(q) = f_{N_m}(1 -  f_{N_m + 1}) G^{(1)}_{N_m}
$$
and
$$
F^{(\Lambda)}(q) = 1 -
\exp\biggl[\frac{\hbar(\Lambda\Omega_c - s q)}{T}\biggr],
$$ 
where $N_m$ in the index of the LL
immediately below the Fermi level,
we
present Eq.~(14) in the following form:
$$
\sigma_{dark}  \propto  
\exp\biggl(-\frac{\hbar \Omega_c}{T}\biggr)\frac{d}{Ldq}\biggl[
\exp\biggl(- \frac{l^2q^2}{2}\biggr)G(q)F(q)\biggr]\biggr|_{q = q^{(1)}}.
$$
\begin{equation}\label{eq15}
=  \overline{G}\biggl(\frac{\hbar s}{TL}\biggr)\biggl(\frac{s}{L\Omega_c}\biggr)
\exp\biggl(- \frac{\hbar\Omega_c}{T}\biggr)
\exp\biggl(- \frac{l^2\Omega_c^2}{2s^2}\biggr).
\end{equation}
We have taken into account that
$Lq^{(1)} =L\Omega_c/s \gg 1$. For example, if $H = 2$~kG,
one obtains $Lq^{1}  \simeq 10$. Hence, for $q \simeq q^{(1)}$ one can set
$G(q) \simeq \overline{G}/Lq$,
where $\overline{G} \simeq 1/2\sqrt{2\pi}N$.  
Similar formula for $\sigma_{dark}$ 
can be derived also for the deformation mechanism
of the electron-phonon interaction.
One can see from Eq.~(15) that the expression for  $\sigma_{dark}$ contains
a small exponential factor $\exp(- \hbar\Omega_c/T)$.
This factor arises because the dark conductivity is associated with
the absorption of acoustic
phonons with the energy $\hbar sq$ close to the LL spacing
by electrons transferring from almost a fully filled LL 
(just below the Fermi level, $N = N_m$),
to the next one which is nearly empty.
However, the number of such photons at $T \ll \hbar\Omega_c$ is  small.
The rate of the processes with the emission of of acoustic
phonons accompanying the electron transitions from an upper LL
is also exponentially small due to a low occupancy of this level and the Pauli
exclusion principle.

Equation~(15) differs from that
calculated previously  by Erukhimov~\cite{17} for
the dark conductivity associated with the accoustic phonon scattering
by the last exponential factor which, according to Eq.~(15), depends on 
the width of the electron localization $l$.
This dependence is attributed to a significant  contribution
of the electron scattering processes with  $q_z \neq 0$ 
which were neglected
in Ref.~17.
As can be seen from comparison
of Eq.~(15) and that obtained in Ref.~17, 
the inclusion of such scattering processes 
results in the replacement of factor $\exp(- L^2\Omega_c^2/2s^2) =
\exp(- \hbar\Omega_c^2/2ms^2)$ (as in ref.~17)
by much larger factor $\exp(- l^2\Omega_c^2/2s^2)$ (as in Eq.~(15)). 
This yields a substantially  higher (exponentially) value of $\sigma_{dark}$ 
than if
the transitions with $q_z \neq 0$ were neglected.
However, the temperature dependences of the dark conductivity
obtained in Ref.~17 and in the present paper coincide.

\section{Photoconductivity}

Using Eq.~(9), we arrive at  
\begin{equation}\label{eq16}
\sigma_{ph} = \sum_{\Lambda}\sigma_{ph}^{(\Lambda)},
\end{equation}
where

$$
\sigma_{ph}^{(\Lambda)}  \propto  
{\cal J}_{\Omega}
\sum_{N}
f_N(1 - f_{N+\Lambda})\int_0^{\infty}
dq\exp\biggl(- \frac{l^2q^2}{2}\biggr)H_N^{(\Lambda)}(q)
$$
$$
\times
[{\cal N}_q\delta^{\prime}(q  + q_{\Omega}^{(\Lambda)})
- ({\cal N}_q + 1)\delta^{\prime}(q  - q_{\Omega}^{(\Lambda)})]
$$
$$
\simeq {\cal J}_{\Omega}\int_0^{\infty}
dq\exp\biggl(- \frac{l^2q^2}{2}\biggr)H^{(\Lambda)}(q)
$$
\begin{equation}\label{eq17}
\times
[{\cal N}_q\delta^{\prime}(q  + q_{\Omega}^{(\Lambda)})
- ({\cal N}_q + 1)\delta^{\prime}(q  - q_{\Omega}^{(\Lambda)})].
\end{equation}
Integrating in the right-hand side of Eq.~(17),
we obtain

\begin{equation}\label{eq18}
\sigma_{ph}^{(\Lambda)}  \propto  - {\cal J}_{\Omega}
\frac{d}{Ldq}\biggl[\exp\biggl(- \frac{l^2q^2}{2}\biggr)
H^{(\Lambda)}(q){\cal N}_q\biggr]\biggr|_{q = - q_{\Omega}^{(\Lambda)}}
\end{equation}
at $\Omega - \Lambda\Omega_c < 0$, and

\begin{equation}\label{19}
\sigma_{ph}^{(\Lambda)}  \propto  {\cal J}_{\Omega}
\frac{d}{Ldq}\biggl[
\exp\biggl(- \frac{l^2q^2}{2}\biggr)H^{(\Lambda)}(q)({\cal N}_q + 1)\biggr]\biggr|_{q = q_{\Omega}^{(\Lambda)}}
\end{equation}
at $\Omega - \Lambda\Omega_c > 0$.
Here,
$$
H^{(\Lambda)}(q) 
\simeq \sum_{N=N_m -\Lambda + 1}^{N_m }f_N(1 - f_{N + \Lambda} )H_N^{(\Lambda)}(q).  
$$

The photoconductivity given by Eqs.~(16) - (19) as a function
of the microwave frequency exhibits pronounced oscillations
in which the photoconductivity sign  alternates.
At the resonances $\Omega = \Lambda\Omega_c$, the photoconductivity 
$\sigma_{ph} = 0$.
Near the resonances $(s/\sqrt{2N_m}L),\, (T/\sqrt{2N_m})
< \hbar|\Omega - \Lambda\Omega_c| < (s/L),\,  (T/\hbar)$,
the value of the photoconductivity varies as a power of 
$(\Omega - \Lambda\Omega_c)$:

\begin{equation}\label{eq20}
\sigma_{ph}^{(\Lambda)} \propto - {\cal J}_{\Omega}\biggl(\frac{TL}{\hbar s}
\biggr)\frac{L^2(\Omega - \Lambda\Omega_c)^2}{s^2} < 0
\end{equation}
at $\Omega - \Lambda\Omega_c < 0$,
and 
\begin{equation}\label{eq21}
\sigma_{ph}^{(\Lambda)} \propto  {\cal J}_{\Omega}\biggl(\frac{TL}{\hbar s}
\biggr)\frac{L^2(\Omega - \Lambda\Omega_c)^2}{s^2} > 0
\end{equation}
when $\Omega - \Lambda\Omega_c > 0$.

Outside the resonances,
i.e., in the ranges 
$(s/L),\, (T/\hbar) <|\Omega - \Lambda\Omega_c|  < \Omega_c$,
 Eqs.~(18) and (19) lead to
\begin{equation}\label{eq22}
\sigma_{ph}^{(\Lambda)} \propto {\cal J}_{\Omega}
\exp\biggl[\frac{\hbar( \Omega - \Lambda\Omega_c)}{T}\biggr]
\exp\biggl[- \frac{l^2(\Omega  - \Lambda\Omega_c)^2}{2s^2}\biggr] > 0
\end{equation}
at $\Omega - \Lambda\Omega_c < 0$, and
\begin{equation}\label{eq23}
\sigma_{ph}^{(\Lambda)} \propto
- \exp\biggl[- \frac{l^2(\Omega - \Lambda\Omega_c)^2}{2s^2}\biggr] < 0
\end{equation}
at $\Omega - \Lambda\Omega_c > 0$.
In this frequency range,
the photoconductivity is determined by two components:
$\sigma_{ph}^{(\Lambda)}$ (which is negative)
and $\sigma_{ph}^{(\Lambda + 1)}$ (whose contribution is positive). As a result,
when $\Lambda\Omega_c < \Omega < (\Lambda + 1)\Omega_c$,
we obtain
\begin{equation}\label{eq24}
\sigma_{ph} \simeq \sigma_{ph}^{(\Lambda)} + \sigma_{ph}^{(\Lambda + 1)},
\end{equation}
so that
$$
\sigma_{ph} \propto {\cal J}_{\Omega}
\biggl\{- \exp\biggl[- \frac{l^2(\Omega - \Lambda\Omega_c)^2}{2s^2}\biggr]
$$
\begin{equation}\label{eq25}
+ \exp\biggl[\frac{\hbar(\Omega - \Lambda\Omega_c - \Omega_c)}{T}\biggr]
\exp\biggl[- \frac{l^2(\Omega - \Lambda\Omega_c  - \Omega_c)^2}{2s^2}\biggr]\biggr\}.
\end{equation}

\section{Discussion}

The negativity of $\sigma_{ph}$ associated
with photon-assisted acoustic scattering near the resonances
at $\Omega \lesssim \Lambda\Omega_c$ and in the ranges
$\Lambda\Omega_c < \Omega (\Lambda + 1)\Omega_c$
is attributed to the following.
When  $\Omega \lesssim \Lambda\Omega_c$, the electron transitions
between LL's contributing to $\sigma_{ph}^{(\Lambda)}$ are 
accompanied by the absorption of acoustic phonons
with the energies $\hbar\omega_q \simeq \Lambda\Omega_c - \Omega$
which are rather small. In this situation, 
the probability of the photon-assisted
phonon absorption decreases with decreasing phonon energy
$\hbar\omega_q \propto q$ because of a decrease in the matrix elements
of the scattering in question with decreasing $q$ (see Fig.~1).
Since the energies of phonons  absorbed near the resonances
are small, the distinction between ${\cal N}_q$ with sufficiently
small and very close
values of $q$ is insignificant.
Due to this, the rate of the absorption of acoustic phonons
with $\hbar\omega_q$ slightly higher than $\hbar(\Lambda\Omega_c - \Omega)$
exceeds that of acoustic phonons
with $\hbar\omega_q \lesssim \hbar(\Lambda\Omega_c - \Omega)$.
In the first case, the 
electron Larmor orbit center displacement $\delta \xi = 
\hbar(\Omega  - \Lambda\Omega_c + \omega_q)/eE > 0$
and therefore the change in the electron potential
energy $\delta \epsilon > 0$, so that
such an act of the electron scattering provides a negative
contribution to the dissipative current, i.e., to ANC.
In contrast, near the resonances but at
$\Omega$ slightly larger than   $\Lambda\Omega_c$,
the scattering acts with  $\delta \xi < 0$ dominate.

Sufficiently far from the resonances
the energies of the emitted (absorbed)
phonons are relatively large. The matrix elements of the scattering
involving such phonons and the phonon number
steeply decrease with increasing
phonon energy (momentum) as shown in Fig.~1.
In this situation, the electron Larmor 
orbit center displacements  
$\delta \xi = \hbar(\Omega  - \Lambda\Omega_c - \omega_q)/eE > 0$ 
corresponding
to the emission of less energetic phonons prevail, that results
in $\sigma_{ph} < 0$ if $\Omega > \Lambda\Omega_c$ but $\Omega$
is still not too close to $(\Lambda + 1)\Omega_c$.
When $\Omega$ increases approaching to and passing   the next resonance,
the situation repeats leading to an oscillative dependence
of   $\sigma_{ph}$ on $\Omega$ if  $\Omega_c$ is kept constant
or on $\Omega_c$ at fixed $\Omega$.
The dependence of the photoconductivity on the frequency
of microwave radiation 
calculated using Eqs.~(16) - (19) for
$L\Omega_s/s = 10$, $l/L = 0.1$,  and different
values of parameter  $b = (\hbar s/TL)$ is shown in Fig.~1.

\begin{figure}
\begin{center}
\includegraphics[width=7.5cm]{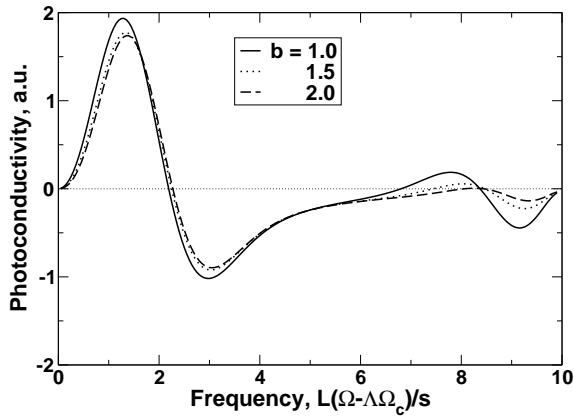}
\end{center}
\label{fig2}
\caption{Photoconductivity vs normalized microwave frequency
in the range $\Lambda\Omega_c \leq \Omega \leq  (\Lambda + 1)\Omega_c$.
}
\end{figure}

As can be drawn from the above formulas at low temperatures
($b \gg 1)$, 
when $\Omega$ increases
in the range $\Lambda\Omega_c \leq \Omega \leq \Lambda\Omega_c$,
 $\sigma_{ph}$ sequentially turns to zero at $\Omega = \Lambda\Omega_c$,
$\Omega = \Lambda\Omega_c + \delta_0$,
and $\Omega = \Lambda\Omega_c + \Delta_0$. 
The photoconductivity becomes
rather small in the range   $ \Lambda\Omega_c + \Delta_0 < \Omega \leq
(\Lambda + 1)\Omega_c $.
The quantity $\delta_0$ can be estimated as 

\begin{equation}\label{eq26}
\delta_0 \simeq \frac{2s}{L}.
\end{equation}
As follows from Eq.~(25), 
\begin{equation}\label{eq27}
\Delta_0  = \Omega_c\biggl[1 - \frac{1}{2(1 + \gamma)}\biggr],
\end{equation}
where $\gamma = (\hbar s/Tl)(s/l\Omega_c)$. One can see from Eq.~(27)
that $\Delta_0/\Omega_c  >  1/2$. The value $\Delta_0$
increases with decreasing temperature. At sufficiently
low temperature ( $T \ll \hbar s/L$),
one obtains $\Delta_0/\Omega_c \lesssim 1$.
For $s = 3\times10^5$~cm/s,
$l = (5 - 10)\times10^{-7}$~cm, $H = 2$~kG, 
and $T = 1$~K, Eqs.~(26) and (27) yield  
$\delta_0/\Omega_c \simeq 1/5$ 
and
$\Delta_0/\Omega_c \simeq 3/5 - 3/4$.
Worth noting the ranges where the calculated photoconductivity
becomes negative  approximately
correspond
to those with zero-resistance~\cite{1,2}.
One needs to point out that the behavior of
the photoconductivity near the resonances can be markedly affected by
the photon-assisted impurity scattering mechanisms and those
associated with other types of
the 2DES disorder~\cite{7,8}. 
The combined contribution of the photon-assisted impurity
and acoustic phonon scattering mechanisms can result in the formation
of the ANC ranges from $\Omega = \Lambda\Omega_c$
to $\Omega = \Lambda\Omega_c + \Delta_0$.

The net dissipative conductivity and its sign are determined
by both $\sigma_{dark}$ and $\sigma_{ph}$.
Different scattering mechanisms can markedly contribute to
the dark component. However, the expression for $\sigma_{dark}$
given by Eq.~(15)
comprises  a small exponential factor $\exp(- \hbar\Omega_c/T)$.
Due to this factor, the photoconductivity
associated with
the  electron-phonon scattering accompanied 
by the absorption of microwave photons
can dominate at rather small values of ${\cal J}_{\Omega}$.
As follows from Eq.~(4),
 the photocurrent
becomes more sensitive to microwave radiation
near the first (cyclotron) resonance owing to
a resonant  increase in ${\cal J}_{\Omega}$.

Both the dark conductivity and photoconductivity
depend to some extent on $|V_{\bf q}|^2$.
In our calculation we assumed that for the piezoelectric accoustic
scattering  $|V_{\bf q}|^2 \propto q^{-1}\exp(- l^2q_z^2/2)$.
One can also use another approximation $|V_{\bf q}|^2 
\propto q^{-1}(1 + l^2q_z^2)^{-2}$
which corresponds to the wave functions of 2D electrons proportional to
$\exp(- |z|/l)$. However, such a change in $|V_{\bf q}|^2$ does not
significantly affect the obtained results.

\section{Conclusion} 

We have calculated the dissipative component of
the dc  conductivity tensor of a 2DES in the transverse magnetic
field and irradiated with microwaves. 
We have demonstrated
that the electron transitions between
the Landau levels stimulated by the absorption
of  microwave photons
accompanied by the emission of acoustic (piezoelectric)
phonons can result in the absolute negative conductivity in rather wide ranges
of the resonance detuning $\Omega - \Lambda\Omega_c$. 
Thus, the ''acoustic'' mechanism of 
the absolute negative conductivity  can contribute
to the formation of the zero-resistance states

\section*{Acknowledgments}

One of the authors (V.R.) is grateful to  V.~Volkov for bringing
Refs.~1 and 2 to his attention and valuable discussions,
and R.~Suris and I.~Aleiner for stimulating comments.
The authors thank A.~Satou for numerical calculations.


\begin{references}



\bibitem{1} 
R.~G.~Mani, J.~H.~Smet, K.~von Klitzing,
V.~Narayanamurti, W.~B.~Johnson, and V.~Umansky,
Nature {\bf 420},  646 (2002).
%
\bibitem{2}
M.~A.~Zudov, R.~R.~Du, L.~N.~Pfeifer, and K.~W.~West,
 Phys.~Rev.~Lett. {\bf 90},  046807-1, (2003).
\bibitem{3}
C.~L.~Yang, M.~A.~Zudov, T.~A.~Knuuttila, R.R.Du, 
 L.~N.~Pfeifer, and K.~W.~West,
arXiv:cond-mat/0303472 (2003).

\bibitem{4} 
P.~W.~Anderson and W.~F.~Brinkman,
arXiv:cond-mat/0302129 (2003).
%
\bibitem{5}
A.~V.~Andreev, I.~L.~Aleiner, and A.~J.~Millis,
arXiv:cond-mat/0302063 (2003).

\bibitem{6}
F.~S.~Bergeret, B.~Huckestein and A~.~F.~Volkov,
arXiv:cond-mat/0303530 (2003).

\bibitem{7}
V.~I.~Ryzhii, Sov.~Phys.-Solid State {\bf 11}, 2078 (1970).
%
\bibitem{8} 
V.~I.~Ryzhii, R.~A.~Suris, and B.~S.~Shchamkhalova,
Sov.~Phys.-Semicond. {\bf 20},  1299 (1986). 


\bibitem{9} 
A.~C.~Durst, S.~Sachdev, N.~Read, and S.~M.~Girvin,
arXiv:cond-mat/0301569 (2003).

\bibitem{10}
V.~I.~Ryzhii,
JETP Lett. {\bf 7}  28 (1968).

\bibitem{11}
V.~I.~Ryzhii, Sov.~Phys.-Solid State {\bf 9}  2286 (1969).

\bibitem{12}
A.~D.~Gladun and V.~I.~Ryzhii,
Sov.~Phys. JETP {\bf 30}, 534 (1970).

%
\bibitem{13} 
V.~Shikin,
JETP Lett. {\bf 77},  236 (2003).

%
\bibitem{14} 
V.~I.~Ryzhii, R.~A.~Suris, and B.~S.~Shchamkhalova,
Sov.~Phys.-Semicond. {\bf 20},  883 (1986). 



\bibitem{15} 
B.~M.~Askerov, {\it Electron Transport Phenomena in Semiconductors}
(World Scientific, Singapore, 1994). 

\bibitem{16}
B.~A.~Tavger and M.~Sh.~Erukhimov,
Sov.~Physics JETP {\bf 24} 354 (1967).

\bibitem{17}
M.~Sh.~Erukhimov,
Sov.~Phys.-Semicond. {\bf 3},  162 (1969).  

\bibitem{18} 
T.~J.~Drammond, W.~Kopp, H.~Morkoc, and M.~Keever,
Appl.~Phys.~Lett. {\bf 41},  277 (1982).

\bibitem{19} 
V.~V.~V'yurkov, A.~D.~Gladun, A.~D.~Malov, and V.~I.~Ryzhii,
Sov.~Phys.- Solid State {\bf 19},  2113 (1977).


\bibitem{20}
L.~S.~Gradsteyn and I.~M.~Ryzhik,
{\it Table of Integrals, Series, and Products,}
(Academic Press, London, 1994).

\end{references}
\end{document}